# Enhancing Phenotype Discovery in Electronic Health Records through Prior Knowledge-Guided Unsupervised Learning

Melanie Mayer, PhD[1,2,3]*; Kimberly Lactaoen, MS[1,2,3]; Gary E. Weissman, MD, MSHP[1,2,3]; Blanca E. Himes, PhD[4]; Rebecca A. Hubbard, PhD[5]

[1] Department of Biostatistics, Epidemiology & Informatics, University of Pennsylvania Perelman School of Medicine, Philadelphia, PA
[2] Pulmonary, Allergy, and Critical Care Medicine Division, University of Pennsylvania Perelman School of Medicine, Philadelphia, PA
[3] Palliative and Advanced Illness Research (PAIR) Center, University of Pennsylvania Perelman School of Medicine, Philadelphia, PA
[4] National Heart, Lung, and Blood Institute (NHLBI), National Institutes of Health, Bethesda, MD
[5] Department of Biostatistics, Brown University, Providence, RI

*Corresponding author: Melanie Mayer, PhD, Department of Biostatistics, Epidemiology & Informatics, University of Pennsylvania, 423 Guardian Dr, Philadelphia, PA (Melanie.Mayer@PennMedicine.upenn.edu)

**Abstract**:

**Objectives:** Unsupervised learning with electronic health record (EHR) data has shown promise for phenotype discovery, but approaches typically disregard existing clinical information, limiting interpretability. We operationalize a Bayesian latent class framework for phenotyping that incorporates domain-specific knowledge to improve clinical meaningfulness of EHR-derived phenotypes and illustrate its utility by identifying an asthma sub-phenotype informed by features of Type 2 (T2) inflammation.

**Materials and methods:** We illustrate a framework for incorporating clinical knowledge into a Bayesian latent class model via informative priors to guide unsupervised clustering toward clinically relevant subgroups. This approach models missingness, accounting for potential missing-not-at-random patterns, and provides patient-level probabilities for phenotype assignment with uncertainty. Using reusable and flexible code, we applied the model to a large asthma EHR cohort, specifying informative priors for T2 inflammation-related features and weakly informative priors for other clinical variables, allowing the data to inform posterior distributions.

**Results:** Using encounter data from January 2017 to February 2024 for 44,642 adult asthma patients, we found a bimodal posterior distribution of phenotype assignment, indicating clear class separation. The T2 inflammation-informed class (38.7%) was characterized by elevated eosinophil levels and allergy markers, plus high healthcare utilization and medication use, despite weakly informative priors on the latter variables.

**Discussion:** Our approach embeds clinical knowledge into unsupervised learning using incomplete EHR data, facilitating data-driven, yet clinically interpretable, phenotype discovery, as shown by the asthma use-case where patterns in the T2-inflammation-informed class suggest an "uncontrolled T2-high" sub-phenotype.

**Conclusion:** Our Bayesian latent class modeling approach supports hypothesis generation and cohort identification in EHR-based studies of heterogeneous diseases without well-established phenotype definitions.

**BACKGROUND AND SIGNIFICANCE**

Uncovering disease subtypes from observable patient characteristics can greatly enhance clinical care by improving disease trajectory prediction and treatment selection, as well as revealing biological mechanisms of disease that may lead to the discovery of precision therapies.[1] Traditional phenotyping approaches rely on rule-based algorithms in which patient phenotype membership is assessed using deterministic criteria defined based on expert clinical knowledge.[2,3] The emergence of large-scale data sources, such as electronic health records (EHR), have provided unprecedented access to clinical information. Consequently, there has been substantial progress in the development of data-driven phenotyping approaches using advanced statistical modeling and machine learning that have improved our ability to characterize diverse populations, phenotype rare diseases, and expand our physiological understanding of disease.[4–6]

The choice of approach to identify phenotypes from data depends in part on whether gold- or silver-standard phenotype labels are available and the proportion of patients with such labels.[7] If a disease is well understood and a gold-standard phenotype can be obtained through procedures such as expert manual chart review, then supervised statistical phenotyping approaches can be applied to improve scalability of phenotype assignment. One can then efficiently apply a validated phenotyping algorithm across large patient populations using routinely collected EHR data, reducing the need for labor-intensive manual review. For complex diseases that are highly variable across patients or not understood well enough to derive gold-standard labels, unsupervised learning approaches can be used for phenotype discovery. In such cases, an understanding of aspects of disease pathophysiology may be helpful to guide the modeling. For example, asthma is likely an umbrella diagnosis for a collection of underlying disease mechanisms, termed endotypes, that lead to similar clinical symptoms and airway inflammation.[8–10] A widely recognized asthma endotype is T-helper type 2 (T2) high asthma, characterized by high levels of T2 inflammation.[11] Patients with T2-high signatures have been shown to have increased responsiveness to inhaled corticosteroid therapy (ICS) and, in the presence of severe disease, to treatment with biologics (e.g., dupilumab, benralizumab, mepolizumab, omalizumab).[12–17] However, T2-high asthma continues to lack a stable and widely accepted definition, responsiveness to biologics is variable, and non-T2-high asthma patients are thought to represent other, less understood endotypes. This heterogeneity underscores asthma as a complex disease for which unsupervised learning methods may yield clinically relevant sub-phenotypes, patient groupings identified from clinical data that may or may not align with biologically defined endotypes.[9]

Prior efforts to guide phenotyping with domain knowledge have included semi-supervised methods, which require subsets of labeled data, and weakly supervised methods, which rely on proxy labels.[18,19] These methods have limited applicability to diseases lacking stable phenotype definitions, as they rely on pre-determined outcome labels. Here, we operationalize a Bayesian latent phenotyping framework that is fully unsupervised but, rather than relying on phenotype labels, clinical knowledge can be embedded at the feature level. By leveraging informative priors, the model steers phenotypes toward known disease characteristics, thereby identifying more clinically meaningful phenotypes.[20] We demonstrate its utility through asthma sub-phenotyping, illustrating how this approach can yield clinically meaningful discovery in a broad, weakly understood condition with implications for future clinical and epidemiological research.

**OBJECTIVE**

We develop an EHR-based phenotype discovery framework that integrates domain knowledge with unsupervised learning using a Bayesian latent class model to improve the clinical interpretability of derived phenotypes for complex diseases lacking gold-standard definitions. We demonstrate the approach by identifying asthma sub-phenotypes informed by features of T2 inflammation.

**MATERIALS AND METHODS**

**Bayesian latent class model**

We consider a setting with patient-level EHR data and implement a Bayesian latent class model, treating the unobserved latent class as the underlying disease sub-phenotype. The latent class, $D$, is modeled conditional on covariates, $X$. Observed data elements – continuous biomarkers and intensity of health care utilization ($Y$), as well as binary clinical codes/medication prescriptions ($W$) – are each modeled conditionally on the latent sub-phenotype. Missingness in $Y$, represented by the indicator $R$, is modeled conditional on $X$ and $D$. Accordingly, the likelihood for the $i^{th}$ patient is given by

$$L(\boldsymbol{\beta}^D, \boldsymbol{\beta}^R, \boldsymbol{\beta}^Y, \boldsymbol{\beta}^W, \boldsymbol{\sigma}^2 | \boldsymbol{X}_i, \boldsymbol{Y}_i, \boldsymbol{W}_i) =$$

$$\sum_{d=0,1} P(D_i = d | \boldsymbol{\beta}^D, \boldsymbol{X}_i) \prod_{j=1}^{J} f(R_{ij} | D_i = d, \boldsymbol{X}_i, \boldsymbol{\beta}_j^R) f(Y_{ij} | D_i = d, \boldsymbol{X}_i, \boldsymbol{\beta}_j^Y, \sigma_j^2)^{R_{ij}} \prod_{k=1}^{K} f(W_{ik} | D_i = d, \boldsymbol{X}_i, \boldsymbol{\beta}_k^W)$$

where $\boldsymbol{\beta}^D$ denotes the parameters quantifying the association between covariates and the underlying patient latent phenotype. Column vectors $\boldsymbol{\beta}^R, \boldsymbol{\beta}^Y, \boldsymbol{\beta}^W$ denote the parameters quantifying the association between the underlying phenotype and (1) the availability of biomarkers and the presence of different measures of health care utilization, (2) continuous biomarkers and rates of health care utilization, and (3) binary clinical codes and medication prescriptions, respectively. For simplicity, in subsequent presentation and application we do not include a dependence on $X$ for $W$ and $Y$. We model associations linearly, such that $D_i \sim Bern(g((1, X_i)\beta^D))$, $R_{ij} \sim Bern\left(g\left((1, X_i, d)\beta_j^R\right)\right)$, $Y_{ij} \sim N((1, d)\beta_j^Y, \sigma_j^2)$, and $W_{ik} \sim Bern((1, d)\beta_k^W)$ for $j = 1, \dots, J$ and $k = 1, \dots, K$, where $g(.) = \exp(.)/(1 + \exp(.))$.

By estimating the parameters within a Bayesian framework, expert knowledge can be incorporated via informative prior distributions to guide latent class estimation towards known characteristics of the disease process. Posterior probabilities of latent class membership, and their credible intervals, serve as patient-level probabilities of sub-phenotype membership and their corresponding uncertainty. The full-information likelihood approach allows one to account for heterogeneity in available data across patients, where each patient's likelihood is computed using only those variables observed for them. If the $j^{th}$ element of $Y$ is missing for individual $i$, $R_{ij} = 0$, and the contribution of $Y_{ij}$ is excluded from the likelihood function. Thus, this model accommodates inconsistently available administrative data, allowing both complete and incomplete data to inform parameter estimation. Many informative EHR-derived data elements, such as diagnostic results or biomarkers, are often missing-not-at-random (MNAR) because their presence depends on the frequency or intensity of a patient's interaction with the healthcare system, which may itself relate to disease status. In accordance with this common missingness pattern found in EHR data, missingness is explicitly modeled as a function of covariates and latent sub-phenotype membership.

To facilitate broad application of this framework, we developed a flexible and publicly available function in R (v4.3.2) implemented using Stan (v2.32.2), a probabilistic programming language commonly used for Bayesian statistical modeling and inference, accessed via the *rstan* user interface.[21–23] Stan models are specified through a joint log-likelihood from which posterior distributions are obtained using the No-U-Turn sampler. Because Stan does not allow direct sampling from discrete parameters, the full joint log-likelihood function was defined and then the binary latent phenotype $D_i$ was marginalized over, see supplemental material for derivation. Source code is available on GitHub (https://github.com/mm4963/prior-guided-EHR-phenotyping).

**EHR study cohort and derived variables**

We illustrate prior construction and model implementation in a retrospective cohort study of adult patients with asthma receiving care at Penn Medicine, a large health care system serving the greater Philadelphia area. Patients were included in the study if they had at least one encounter between January 1, 2017, and February 29, 2024 which was accompanied by (1) an International Classification of Diseases, 10th revision (ICD-10) diagnosis code indicating asthma (J45*) and (2) a prescription for a short-acting $\beta_2$-agonist (SABA) medication, not necessarily from the same encounter. We further restricted our analysis to adult patients with sufficient data to support phenotyping by including only encounter data for individuals aged 18 years or older with at least one year between their first and last included encounter. The University of Pennsylvania Institutional Review Board approved our study under protocol number 824789. A waiver of HIPAA Authorization was granted for the conduct of this research thus formal consent was not obtained.

Patient covariates ($X$) included sex (male/female), age at first encounter (years), race (Black, White, or other), ethnicity (Hispanic/Latino or not Hispanic/Latino), insurance (Medicaid, Medicare, or private), smoking status (never, ever, or current smoker), and body mass index (BMI) (kg/m$^2$).

Continuous variables ($Y$) included multiple biomarkers and healthcare utilization measures. Biomarkers included the last recorded total immunoglobulin E (IgE) test result (kU/L), polygenic risk scores for asthma computed from coefficients based on a genome-wide association study conducted within UK Biobank (SCORE)[24–26], pulmonary function test (PFT) computed as last recorded ratio of Forced Expiratory Volume in one second to the Forced Vital Capacity (FEV1/FVC), and average blood eosinophil count (EOS) (10$^3$ cells/$\mu L$). Healthcare utilization variables quantified annual rates of asthma encounters (encounters with an ICD-10 code for asthma), asthma exacerbations (encounters with a primary ICD-10 code for asthma and a prescription for an OCS medication), short-acting beta agonist (SABA) prescriptions, and emergency visits (ED) (ED encounters with an ICD-10 code for asthma). All continuous variables were standardized (mean = 0, SD = 1), and those exhibiting skewed distributions were log-transformed. For health care utilization variables, patients with no encounters were treated as missing, allowing us to incorporate both the intensity of utilization among those with encounters as well as the probability of engaging with the health care system in that manner.

For clinical codes ($W$), we included a composite indicator for allergies, defined as whether the patient had any encounter with an ICD-10 code whose description included the text "allerg" or "atopic" (see supplemental material for more details), as well as whether the patient ever had a positive allergy skin prick test, an ICS prescription, or a prescription for ICS and a long-acting beta-agonist (ICS/LABA) combination medication.

**Specification of informative priors**

Expert-guided deterministic classification based on presence or absence of certain clinical conditions or biomarkers falling in a particular range may be effective for disease diagnosis in the context of clinical care. However, as noted, required data elements may not be available in secondary analysis of EHR data as not all patients receive the necessary diagnostic evaluation. Nonetheless, we can encode expert knowledge represented by these deterministic rules by, for example, assigning informative priors that, given the low prevalence of the characteristic, favor it being absent in individuals without the disease such that there are very few false positives. Conceptually, this corresponds to assigning priors that nudge the feature towards having high specificity for identifying T2-high asthma. Depending on the hypothesized prevalence of the latent phenotype relative to the characteristic, priors can also encode reduced sensitivity to reflect that the feature may not be present in all individuals with T2-high asthma.

Consider Table 1a, a contingency table relating the predictive ability of the dichotomization of a continuous biomarker $Y_j$ at a threshold $c$ to classify patients into latent phenotype class, $D$. In practice, we have empirical values for $n_{R_j=0,Y_j=0}$, $n_{R_j=0,Y_j=1}$, and $n_{R_j=0}$, such that the row margins of Table 1a are fixed. Here, $R_j = 0$ indicates $Y_j$ is observed. Thus, one can specify any two of the following: sensitivity, $P(Y_j \geq c | D = 1, R_j = 0)$; specificity, $P(Y_j < c | D = 0, R_j = 0)$; and prevalence of the phenotype amongst those without missingness, $P(D = 1 | R_j = 0)$. Due to the fixed margins, the third is fixed given the other two, and setting informative priors for parameter values that correspond to a target sensitivity and specificity substantially impacts the resulting posterior probability for the prevalence of the latent class. The target sensitivity and specificity therefore should be chosen with care.

**Table 1.** Contingency tables illustrating the classification of patients into latent class membership, $D$.

(a) Conceptual contingency table based on the dichotomization of a continuous biomarker, $Y_j$, at threshold c for patients with $Y_j$ observed, such that missingness indicator $R_j = 0$.

|  | BIOMARKER THRESHOLD NOT REACHED $Y_j < c$ | BIOMARKER THRESHOLD REACHED $Y_j \geq c$ |  |
|---|---|---|---|
| **BELONGS TO LATENT CLASS** $(D = 1)$ | $\Sigma_{n_{R_j=0}} I\{Y_j < c, D = 1\}$ <br><br> False Negative (FN) | $\Sigma_{n_{R_j=0}} I\{Y_j \geq c, D = 1\}$ <br><br> True Positive (TP) | $n_{R_j=0,D=1} = \Sigma_N I\{D = 1, R_j = 0\}$ |
| **DOES NOT BELONG TO LATENT CLASS** $(D = 0)$ | $\Sigma_{n_{R_j=0}} I\{Y_j < c, D = 0\}$ <br><br> True Negative (TN) | $\Sigma_{n_{R_j=0}} I\{Y_j \geq c, D = 0\}$ <br><br> False Positive (FP) | $n_{R_j=0,D=0} = \Sigma_N I\{D = 0, R_j = 0\}$ |
|  | $n_{R_j=0,Y_j=0} = \Sigma_N I\{Y_j < c,, R_j = 0\}$ | $n_{R_j=0,Y_j=1} = \Sigma_N I\{Y_j \geq c,, R_j = 0\}$ | $n_{R_j=0}$ |

(b) Example contingency table showing the use of average blood eosinophil count ($Y_{EOS}$) dichotomized at $0.15 \times 10^3 \; cells/\mu L$ to indicate membership to the latent T2-high informed asthma sub-phenotype for the subgroup of patients with observed blood eosinophil counts. Values represent percentages rather than sample sizes and reflect the subset to patients with measured blood eosinophil count. Grey cells indicate empirically fixed values.

|  | $Y_{EOS} < 0.15$ | $Y_{EOS} \geq 0.15$ |  |
|---|---|---|---|
| $D = 1$ | 14.1 | 42.3 | 56.4 |
| $D = 0$ | 33.0 | 10.6 | 43.6 |
|  | 47.1 | 52.8 | 100 |

For the asthma application, informative priors for certain model parameters, specifically those relating to variables known to be associated with T2 inflammation, were chosen in an effort to guide the latent variable towards characterizing a T2-high asthma sub-phenotype. Allergic and eosinophilic asthma are mechanistically driven by type 2 inflammation and thus are considered subsets within the T2-high asthma sub-phenotype.[11,27,28] Thus, we applied informative priors for the parameters associated to EOS, positive allergy skin prick test, and the composite allergy indicator. Weakly informative priors were used for all other parameters. While we could have applied informative priors to other variables as well, such as IgE, we chose not to due to the inconclusiveness for sub-phenotyping T2-high asthmas in past studies.[29,30] Below, we illustrate the process of specifying informative priors using measured blood eosinophil count. Details on how the additional informative priors were derived using prior knowledge, along with the full specification of all priors used in the analysis, are provided in the supplemental material.

Blood eosinophil count, denoted $Y_{EOS}$, is a biomarker used as an indicator of T2-high asthma in clinical practice. Certain current guidelines recommend labeling patients with $Y_{EOS} \geq 0.15 \times 10^3$ cells/$\mu L$ as T2-high.[12,31] In our cohort, 75.8% of our cohort have observed $Y_{EOS}$, and of this subgroup $P(Y_{EOS} \geq 0.15 | R_{EOS} = 0) = 0.53$ and $P(Y_{EOS} < 0.15 | R_{EOS} = 0) = 0.47$, empirically. Some asthma medications, including ICS and biologics, have been shown to reduce the quantity of blood eosinophils.[32] Since we are considering the average blood eosinophil count across all measures during follow-up, which does not account for the temporal order of when the biomarker was measured relative to receiving medication, we would expect $Y_{EOS} \geq 0.15 \times 10^3$ cells/$\mu L$ to be a good indicator of $D = 1$ but $Y_{EOS} < 0.15 \times 10^3$ cells/$\mu L$ to be less reliable for $D = 0$. Suppose we believe that around 80% of patients with observed $Y_{EOS} \geq 0.15 \times 10^3$ cells/$\mu L$ fall into $D = 1$ and 70% of patients with observed $Y_{EOS} < 0.15 \times 10^3$ cells/$\mu L$ fall into $D = 0$. Table 1b shows how every cell of the contingency table would be filled using these proportions in our cohort. The resulting target sensitivity would be $P(Y_{EOS} \geq 0.15 | D = 1, R_{EOS} = 0) = \frac{42.3}{56.5} = 0.75$, specificity of $P(Y_{EOS} < 0.15 | D = 0, R_{EOS} = 0) = \frac{33.0}{43.6} = 0.76$, and prevalence of the latent class of 56.4% amongst the subgroup with observed $Y_{EOS}$. This is a reasonable target given that, though there is no consensus on the exact prevalence of T2-high asthma amongst asthmatic patients, it is thought to be around 50%.[33–35]

To embed these targets into our priors, we consider how we are modeling $Y_{EOS}$, which is log transformed and then standardized before implementing our model, such that $\frac{\log Y_{EOS} - \mu_{\log Y_{EOS}}}{\sigma_{\log Y_{EOS}}} \sim Normal(\beta_0^{EOS} + \beta_1^{EOS}D, 1)$, where $\mu_{\log Y_{EOS}} = \frac{1}{n}\Sigma \log(Y_{EOS,i})$ and $\sigma_{\log Y_{EOS}} = \sqrt{\frac{1}{n-1}\Sigma(\log Y_{EOS,i} - \mu_{\log Y_{EOS}})^2}$. It follows that

$$P\left(\frac{\log Y_{EOS} - \mu_{\log Y_{EOS}}}{\sigma_{\log Y_{EOS}}} < \frac{\log(0.15) - \mu_{\log Y_{EOS}}}{\sigma_{\log Y_{EOS}}} \middle| D, R_j = 0\right) = \Phi\left(\frac{\log(0.15) - \mu_{\log Y_{EOS}}}{\sigma_{\log Y_{EOS}}} - \beta_0^{EOS} - \beta_1^{EOS}D\right),$$

where $\Phi(.)$ is the cumulative distribution function of the standard normal distribution. We can use this formulation to select priors for sensitivity, where $P(Y_{EOS} < 0.15 | D = 0, R_j = 0) = \Phi\left(\frac{\log(0.15) - \mu_{\log Y_{EOS}}}{\sigma_{\log Y_{EOS}}} - \beta_0^{EOS}\right)$, a function of the parameter $\beta_0^{EOS}$. We implement the priors $\beta_0^{EOS} \sim N(-0.68, 0.05)$ and $\beta_1^{EOS} \sim N(0.70 - \beta_0^{EOS}, 0.05)$, which are centered at the target sensitivity of 75% and specificity of 76%. Note that this choice is in line with the belief that $Y_{EOS}$ is higher on average for those with T2-high asthma ($D = 1$) compared to those with non-T2-high asthma ($D = 0$).

We ran the model over 4 chains, discarding the first 200 iterations of the MCMC sampler and based all results on a subsequent sample of 1000 draws per chain from the posterior distribution of the model parameters. Convergence was assessed by visually inspecting trace plots and ensuring the R hat statistic was less than 1.1 for all parameters.[36]

## RESULTS

We developed an open-source R function to readily implement the prior-informed latent class framework. To demonstrate this workflow, we applied this approach to the adult asthma EHR cohort. Visual inspection of trace plots and R-hat statistics indicated satisfactory convergence for all parameters (see supplemental material). Below, we first summarize cohort characteristics to contextualize the application, followed by results of the Bayesian latent class analysis.

### Asthma cohort population

Data from the Penn Medicine cohort included 46,546 adults with asthma. Participants were excluded due to missingness in covariates, resulting in 44,642 patients. A summary of the baseline demographic characteristics for our cohort are provided in Table 2. Our cohort of patients with asthma included a high proportion of Black (n = 15,421, 34.5%) and female (n = 31,581, 69.6%) patients. Notably, this is a higher proportion of Black patients and women then in the general population of patients receiving care in the University of Pennsylvania Health System.[37] Our cohort also had a high average BMI of 31.4, which falls within the obese range.

Table 2. Characteristics of study population of adult asthma patients stratified according to posterior probability of belonging to the latent T2-high asthma sub-phenotype (D = 1)

|  | Total | P(D = 1) < 0.5 | P(D = 1) ≥ 0.5 |
|---|---|---|---|
|  | N = 44642 (100%) | n = 28424 (63.7%) | n = 16218 (36.3%) |
| **Covariates** | | | |
| **Sex = Male (n (%))** | 13061 (29.3) | 8633 (30.4) | 4428 (27.3) |
| **Age at 1st encounter (mean (SD))** | 47.38 (17.48) | 46.37 (17.72) | 49.16 (16.90) |
| **Race (n (%))** | | | |
|   White | 24033 (53.8) | 15560 (54.7) | 8473 (52.2) |
|   Black | 15421 (34.5) | 9759 (34.3) | 5662 (34.9) |
|   Other | 24033 (11.7) | 3105 (11.0) | 2083 (12.9) |
| **Ethnicity = Hispanic/Latino (n (%))** | 2412 (5.4) | 1426 (5.0) | 986 (6.1) |
| **Insurance Type (n (%))** | | | |
|   Medicaid | 7608 (17.0) | 4675 (16.4) | 2933 (18.1) |
|   Medicare | 12293 (27.5) | 7186 (25.3) | 5107 (31.5) |
|   Private | 24741 (55.4) | 16563 (58.3) | 8178 (50.4) |
| **Smoking Status (n (%))** | | | |
|   Current Smoker | 4055 (9.1) | 2899 (10.2) | 1156 (7.1) |
|   Ever Smoker | 12761 (28.6) | 8038 (28.3) | 4723 (29.1) |
|   Never Smoked | 27826 (62.3) | 17487 (61.5) | 10339 (63.8) |
| **BMI (mean (SD))** | 31.36 (8.33) | 30.90 (8.07) | 32.17 (8.71) |
| **Variables Without Missingness** | | | |
| **Rate of asthma encounters per year (mean (SD))** | 2.62 (2.35) | 1.80 (1.24) | 4.06 (3.05) |
| **Rate of SABA prescriptions per year (mean (SD))** | 2.20 (2.96) | 1.67 (2.01) | 3.12 (3.96) |
| **Received an ICS prescription (n (%))** | 16150 (36.2) | 7993 (28.1) | 8157 (50.3) |
| **Received an ICS/LABA prescription (n (%)** | 25895 (58.0) | 11971 (42.1) | 13924 (85.9) |
| **Positive skin prick allergy test (n (%))** | 688 (1.5) | 40 (0.1) | 648 (4.0) |
| **Allergic ICD-10 code (n (%))** | 16802 (37.6) | 6657 (23.4) | 10145 (62.6) |
| **Variables With Missingness (mean (SD))** | | | |
| **Rate of ED visits associated with asthma per year** | 0.10 (0.49) | 0.05 (0.22) | 0.18 (0.75) |
| **Rate of asthma exacerbations per year** | 0.57 (0.65) | 0.33 (0.20) | 0.66 (0.73) |
| **Total immunoglobulin E (IgE) (kU/L)** | 240.13 (415.82) | 98.73 (165.64) | 245.18 (421.16) |
| **Polygenic risk scores for asthma ($\times 10^{-3}$)** | -2.22 (1.98) | -2.26 (1.92) | -2.20 (2.02) |
| **FEV1/FVC** | 0.75 (0.12) | 0.75 (0.12) | 0.75 (0.11) |
| **Blood eosinophil count ($10^3$ cells/$\mu L$)** | 0.20 (0.37) | 0.18 (0.45) | 0.22 (0.21) |
| **Missingness Indicators (n (%))** | | | |
| **No ED visit associated with asthma** | 38056 (85.2) | 25502 (89.7) | 12554 (77.4) |
| **No asthma exacerbation** | 32570 (73.0) | 25153 (88.5) | 7417 (45.7) |
| **Total immunoglobulin E (IgE) test missing** | 40989 (91.8) | 28298 (99.6) | 12691 (78.3) |
| **Polygenic risk scores missing** | 43809 (98.1) | 28103 (98.9) | 15706 (96.8) |
| **FEV1/FVC test missing** | 36293 (81.3) | 25925 (91.2) | 10368 (63.9) |
| **Blood eosinophil count missing** | 10805 (24.2) | 8180 (28.8) | 2625 (16.2) |

BMI: body mass index; ED: emergency department; FEV1/FVC: Forced Expiratory Volume in one second to the Forced Vital Capacity; ICD-10: International Classification of Diseases, 10th revision; ICS: inhaled corticosteroid; LABA: long-acting $\beta_2$-agonist; SABA: short-acting $\beta_2$-agonist; SD: standard deviation

All continuous variables aside from the polygenic risk score were found to be skewed and thus were log transformed prior to standardization. The cohort included 688 patients (1.5%) who had a positive allergic skin prick test and 16,802 patients (37.6%) who had an ICD-10 code whose description includes the phrase "allerg" or "atopic". Overall, 33, 837 patients (75.8%) had at least one record for EOS, while only 3,653 (8.2%) had at least one measurement for total IgE. As stated above and in Table 1b, of those patients with measured EOS, 17,876 patients (52.8%) had an average EOS above $0.15 \times 10^3$ cells/$\mu L$.

**Latent class results**

The empirical distribution of the posterior probabilities of belonging to the T2-high asthma sub-phenotype in this sample can be seen in Figure 1. We found a bimodal distribution, with two peaks around 0.03 and 0.98, indicating a clear separation in classification based on the latent variable. There are many patients with either very low or high posterior probabilities of having T2-high asthma. Using a cutoff of 0.5 for the posterior probability, 36.3% of patients were labeled as T2-high.

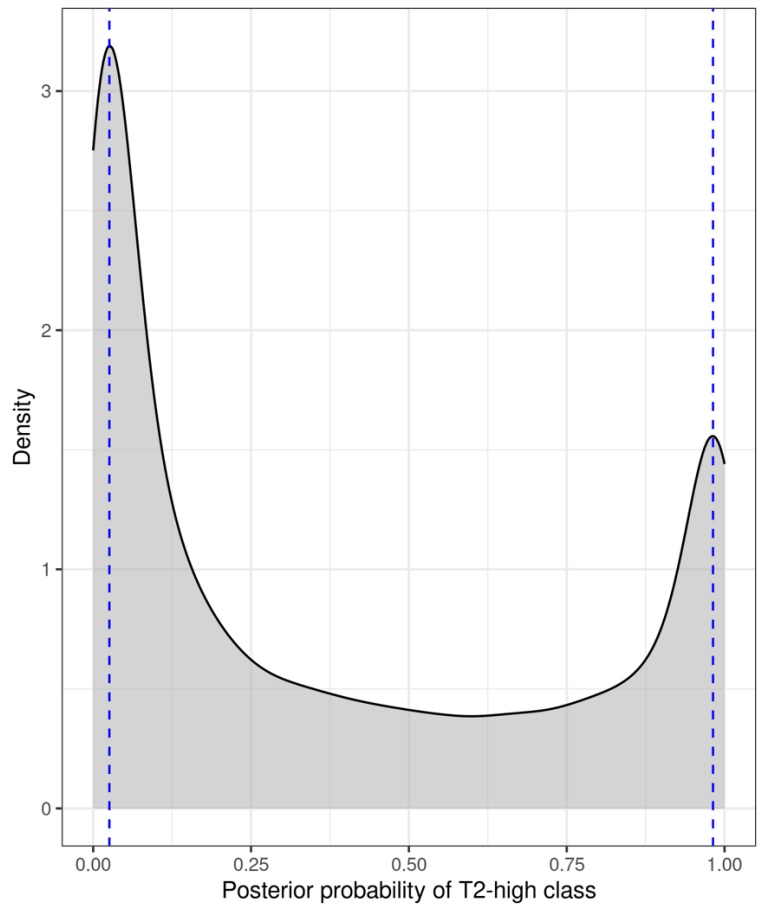

*Figure 1.* *Density of the posterior probability of membership into the T2-high latent class as estimated by the Bayesian latent class model. Dashed lines indicate two modal peaks at 0.03 and 0.98.*

For comparison, we implemented a simple rule-based phenotyping approach using three indicators: allergic skin prick test (positive/negative/missing), blood eosinophils ⩾0.15×10³/μL (positive/negative/missing), and presence of an allergic ICD code (present/absent). Rule-based T2-high was assigned "Positive" if any indicator was positive/present; "Negative" if at least one of allergic skin prick test or eosinophils was observed negative and the allergic ICD code was absent; otherwise "Indeterminate". Under this rule, 22.9% (n=10,218) of our asthma cohort were classified as Negative, 62.4% (n=27,852) as Positive, and 14.7% (n=6,572) as Indeterminate. Among rule-based Negatives, 78.7% were categorized as non-T2-high by our model, while 47.9% of rule-based Positives were categorized as T2-high. Among Indeterminate cases, 89.5% were classified as non-T2-high.

Table 2 summarizes the characteristics of adult asthma patients stratified by this classification. We found little difference in their polygenic risk scores for asthma or pulmonary function test results (FEV1/FVC) compared to those below this threshold, two indicators of general asthma that are not necessarily specific to the T2-high sub-phenotype. However, those labeled T2-high had higher levels of health care utilization, were more likely to receive medication prescriptions, and had lower levels of missingness. For example, they had an average of 4.1 asthma encounters per year (vs. 1.8), 85.9% received an ICS/LABA prescription (vs. 42.1%), and 78.3% were missing IgE test results (vs. 99.6%). Despite using weakly informative priors for parameters related to IgE, a biomarker previously proposed as an indicator of the sub-phenotype, patients classified as T2-high had a higher average total IgE level (246.5 kU/L) compared to those not classified as T2-high (104.2 kU/L).

To further examine how individual variables contributed to sub-phenotype assignment, Figure 2 shows the distribution of expected posterior probability of latent class membership stratified by selected variables. For allergic ICD-10 codes and EOS stratified at $0.15 \times 10^3$ cells/$\mu L$, two variables for which informative priors were used, patients with these characteristics had distributions with higher concentrations at higher posterior probabilities of belonging to the latent class compared to those without. For example, the average posterior probability of belonging to the latent class was 0.60 (0.59, 0.61) among those with an allergic ICD-10 code, compared to 0.26 (0.25, 0.27) among those without. Figure 2 also shows the posterior probability density stratified by smoking status and ICS/LABA prescription. Although weakly informative priors were placed on these variables, a similar trend was observed. For instance, the average posterior probability of belonging to the latent class was 0.32 for current smokers and 0.39 for both ever and never smokers.

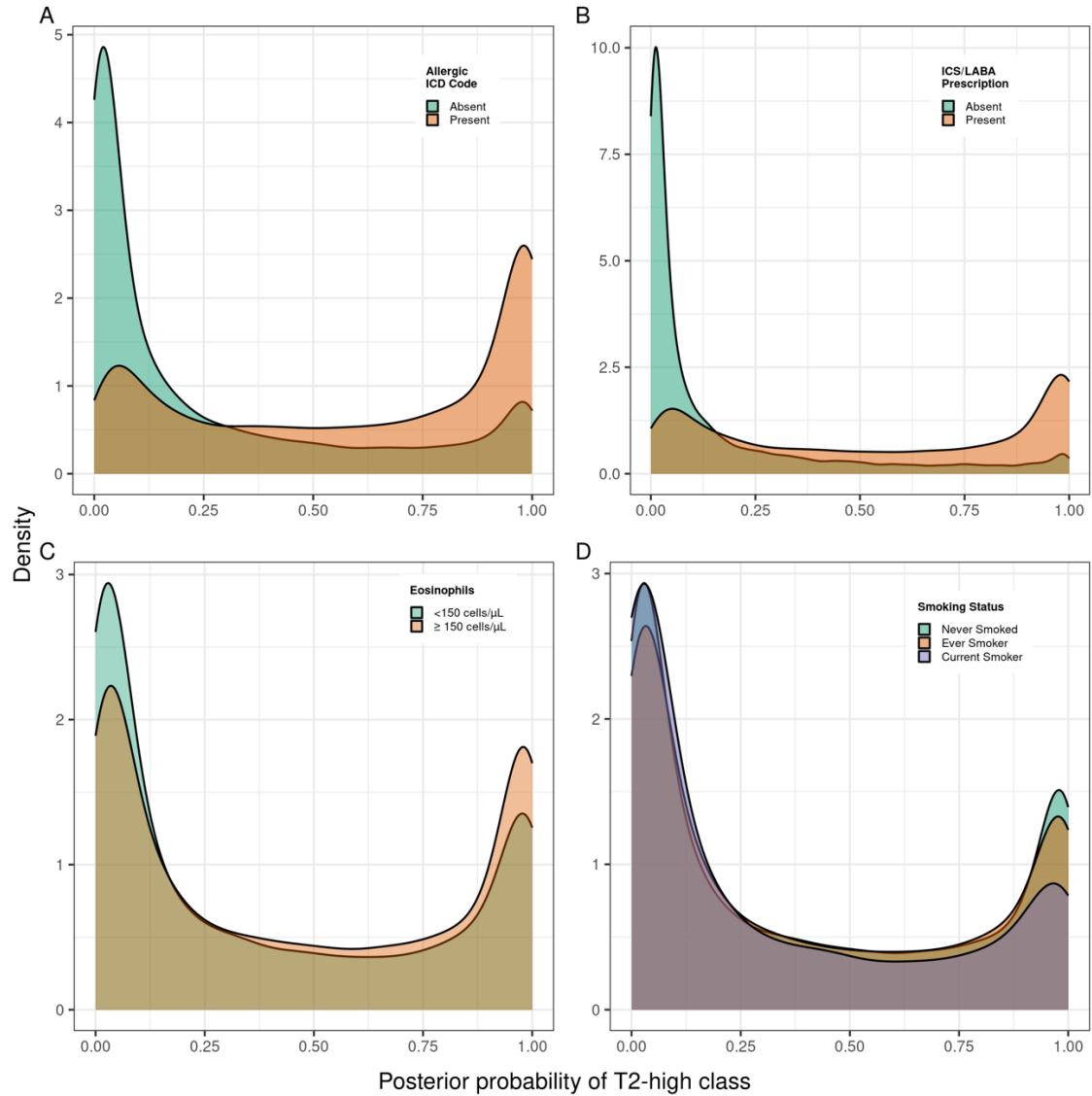

*Figure 2.* Density of the posterior probability of membership into the T2-high latent class by features stratified by clinical features. Panels show (A) allergic ICD code (absent/present), (B) inhaled corticosteroid therapy and a long-acting beta-agonist (ICS/LABA) combination medication prescription (absent/present), (C) average blood eosinophil count (<150 cells/$\mu L$ vs $\geq cells/\mu L$), and (D) smoking status (never/ever/current).

## DISCUSSION

This study presents a framework that embeds prior knowledge into an unsupervised learning approach for disease sub-phenotyping using EHR data. We demonstrated its utility by applying it to a well-studied but inconsistently defined asthma sub-phenotype, T2-high asthma, to categorize a large, diverse adult asthma cohort. Our Bayesian latent class model maximized the amount of information used from variables passively and inconsistently collected through routine clinical care to estimate posterior probabilities of belonging to the latent sub-phenotype. Informative priors on relevant parameters reflected known discriminative markers of T2-high asthma, guiding clustering toward a targeted, biologically meaningful subgroup.

The posterior probabilities of belonging to the latent class displayed a bimodal distribution, with modes at high and low probabilities, indicating strong separation of patients into two clusters. While general asthma indicators (e.g., FEV1/FVC, polygenic risk scores) were similarly distributed across latent classes, membership in the T2-high class was characterized by elevated markers of Type 2 inflammation, such as eosinophil count and allergy indicators, as well as higher healthcare utilization and medication use, despite weakly informative priors on these latter features. This pattern suggests the latent class represents "uncontrolled T2-high" asthma. Its lower prevalence (38.7%) compared to the hypothesized ~50% prevalence of T2-high asthma amongst asthma patients aligns with the reality that not all T2-high patients are uncontrolled.[11] Such a subgroup may be more relevant for clinical decision-making, for instance in identifying patients eligible for biologic therapies.

This interpretation also helps explain the partial concordance with the rule-based comparator, where many rule-based Positives did not meet the probabilistic threshold for T2-high. Because our rule set was hypothesis-driven rather than discovery-oriented, the two approaches may target different subgroups. Prior work has used expert-defined heuristics to deterministically classify patients as T2-high.[38,39] Although appropriate for diagnosis in clinical care, this approach is difficult to implement in secondary analysis of EHR data where researchers lack control over obtaining the necessary diagnostic evaluations. This leads, as in our comparator, to many patients being classified as Indeterminate. In contrast, our model yields probabilistic assignments for all patients, even when key EHR data elements are missing. Moreover, deterministic rules are subjective, fail to account for uncertainty in phenotype assignment, and are suboptimal for phenotype discovery. [40]

The absence of gold- or silver-standard labels in EHR data has driven a shift toward unsupervised learning.[41,42] However, without the incorporation of prior knowledge, unsupervised clustering studies rely on post hoc interpretation of emergent patterns in unlabeled data.[43-47] By incorporating clinical knowledge directly into clustering through informative priors, our framework bridges the traditional expert-defined rules approach and purely data-driven discovery.[2,48-50] Investigators can specify priors ranging from strongly to weakly informative depending on the available evidence, allowing flexibility between the influence of prior knowledge and observed data. In our asthma application, constraining selected parameters based on the understood discriminative performance of key markers steered clustering toward a clinically meaningful T2-high subgroup. The resulting clusters roughly distinguished, amongst other features, patients with higher eosinophil counts, allergy-related markers, and greater healthcare utilization, from those without. Our approach enables direct, theory-informed interpretation while still supporting data-driven discovery.

In addition, our Bayesian framework addresses common challenges in EHR-based studies, including incomplete data with informative missingness, such as missing-not-at-random patterns where data availability is linked to disease severity. For example, immunoglobulin E testing was missing for 81.3% of the cohort. Our model did not exclude these patients and modeled this missingness as dependent on covariates as well as the latent class, showing it was missing for 78.3% of patients classified as T2-high versus 99.6% of those not classified as T2-high, indicating a strong association without deterministically defining class membership. Beyond handling missing data, the model produces probabilistic rather than hard class assignments. As shown in Figure 1, most patients had posterior probabilities near the extremes, though many were closer to 50%. Together with the uncertainty quantification easily derived from each patient's full posterior distribution, these features further enhance interpretability and support downstream analyses.

It is important to recognize that EHR data are collected primarily for clinical care and administrative purposes rather than research, which limits causal interpretations and phenotype validation.[51–53] For example, while atopy increases the likelihood of belonging to the T2-high sub-phenotype, a positive skin prick test or elevated eosinophil count does not confirm that an allergen is causing asthma symptoms. Our probabilistic approach accommodates some uncertainty but cannot fully overcome the inherent limitations of EHR-derived data. Consequently, this framework is best suited for phenotype exploration and hypothesis generation rather than individual-level risk prediction. Posterior probabilities can help identify homogeneous subpopulations for further investigation, such as targeted enrollment in clinical trials, rather than serving as direct predictors of clinical outcomes.

**CONCLUSION**

We present a flexible and interpretable Bayesian latent class modeling framework that embeds prior knowledge to enhance unsupervised disease sub-phenotyping. Applied to a diverse adult asthma cohort generated from EHR data, this approach enabled theory-driven clustering aligned with the T2-high asthma sub-phenotype and revealed a clinically relevant subgroup suggestive of "uncontrolled T2-high" asthma, distinguished not only by biomarkers of type 2 inflammation but also by high healthcare utilization and medication use. By embedding informative priors, handling incomplete and non-randomly missing data, and yielding probabilistic class assignments, our method addresses key limitations of both rule-based and traditional unsupervised phenotyping approaches for EHR data. This framework supports phenotype discovery and hypothesis generation and is broadly applicable to heterogeneous diseases without established phenotype definitions using complex, incomplete EHR-derived data.

**Author Contributions:** RH, BH, and GW were responsible for funding acquisition. RH, BH, GW, and MM conceptualized the study. BH and KL performed data curation. KL performed EHR data processing. RH and MM developed the methodology. MM developed software code, performed the formal analyses, and created visualizations. RH and MM wrote the original draft, and all authors revised and approved the final manuscript.

**Acknowledgements:** We would like to thank Sunil Thomas, MBA/TM from the University of Pennsylvania Penn Data Store for extracting the electronic health record data, Alex Moffett, MD for providing pre-processed pulmonary function test data, and John Gregg, PhD for calculating polygenic risk scores used for this project. We would also like to thank Patrick Gleeson, MD, MSCE for his expert clinical assistance in the evaluation of electronic health records. We acknowledge the Penn Medicine BioBank (PMBB) for providing data and thank the patient- participants of Penn Medicine who consented to participate in this research program. We would also like to thank the Penn Medicine BioBank team and Regeneron Genetics Center for providing genetic variant data for analysis. The PMBB is approved under IRB protocol# 813913 and supported by Perelman School of Medicine at University of Pennsylvania, a gift from the Smilow family, and the National Center for Advancing Translational Sciences of the National Institutes of Health under CTSA award number UL1TR001878.

**Sources of funding**: Research reported in this publication was supported by the National Institutes of Health (NIH) National Heart, Lung, And Blood Institute (NHLBI) under Award Number R01HL162354. The funders had no role in study design, data collection and analysis, decision to publish, or preparation of the manuscript. The content is solely the responsibility of the authors and does not necessarily represent the official views of the National Institutes of Health or other funding agencies.

**Conflicts of Interest**: Dr. Himes contributed to this article as an employee of the University of Pennsylvania. The views expressed are her own and do not necessarily represent the views of the National Institutes of Health or the United States Government.

**Data Availability Statements:** Based on ethical and legal considerations, such as that the data falls under HIPAA regulations and was not collected with informed consent, the University of Pennsylvania Institutional Review Board has determined that the electronic health record (EHR) data used in this study cannot be shared widely. Upon reasonable request to the authors, deidentified EHR data may be provided.

## Supplemental Material

### Section S1. Additional EHR Data Processing Details

The composite indicator for allergies was derived from the list of ICD-10 codes specified in Table S2, all of whose descriptions contain the text "allerg" or "atopic". This indicator was coded as present if a patient had at least one encounter associated with any of these codes. ICD-10 code data in our study were limited to codes recorded at encounters with a diagnosis code for asthma (J45*) or COPD (J44*).

Pulmonary function tests (PFTs) were obtained from a separate data extraction team at Penn Medicine, in which spirometry results were collected from select pulmonary diagnostic labs within the University of Pennsylvania Health System. This data included results from February 2000 to April 2024 and were subsequently linked to our study cohort. The ratio of the Forced Expiratory Volume in one second to the Forced Vital Capacity (FEV1/FVC) was used for analysis. We used the most recent pre-bronchodilator spirometry measurements when available. For patients without a bronchodilator test, we used their most recent spirometry values of any type.

Polygenic risk scores were computed using coefficients from the UK Biobank and genetic information from Penn Medicine Biobank (PMBB) participants. The PMBB is a voluntary research program in which patients consent to provide biospecimens and link their electronic health record data for research use. Genetic data from PMBB participants were linked to our study cohort.

*Table S1. ICD-10 codes included in allergy indicator condition*

| ICD-10 Code | Description |
|---|---|
| J309 | Allergic rhinitis unspecified |
| J301 | Allergic rhinitis due to pollen |
| J3081 | Allergic rhinitis due to animal cat dog hair and dander |
| Z9109 | Other allergy status other than to drugs and biological substances |
| J3089 | Other allergic rhinitis |
| Z91012 | Allergy to eggs |
| Z91014 | Allergy to mammalian meats |
| Z91018 | Allergy to other foods |
| Z91013 | Allergy to seafood |
| Z91040 | Latex allergy status |
| Z91038 | Other insect allergy status |
| Z91030 | Bee allergy status |
| Z91010 | Allergy to peanuts |
| Z0182 | Encounter for allergy testing |
| J305 | Allergic rhinitis due to food |
| Z91048 | Other nonmedicinal substance allergy status |

| Z91041 | Radiographic dye allergy status |
| Z91011 | Allergy to milk products |
| Z9102 | Food additives allergy status |
| Z88 | Allergy status to drugs medicaments and biological substances |
| L23 | Allergic contact dermatitis |
| L20 | Atopic dermatitis |

**Section S3. Marginalized log-likelihood derivation**

Stan does not allow direct sampling from discrete parameters, requiring marginalization of discrete parameters. In our context, the latent phenotype, $D_i$, constitutes a binary parameter. Here we defined the full joint log-likelihood function, marginalized over $D_i$, needed to implement our model in Stan. The likelihood for the $i$th patient is:

$$L(\boldsymbol{\beta}^D, \boldsymbol{\beta}^R, \boldsymbol{\beta}^Y, \boldsymbol{\beta}^W, \sigma^2 | \boldsymbol{X}_i, \boldsymbol{Y}_i, \boldsymbol{W}_i) =$$

$$\sum_{d=0,1} P(D_i = d | \boldsymbol{\beta}^D, \boldsymbol{X}_i) \prod_{j=1}^{J} f(R_{ij} | D_i = d, \boldsymbol{X}_i, \boldsymbol{\beta}_j^R) f(Y_{ij} | D_i = d, \boldsymbol{X}_i, \boldsymbol{\beta}_j^Y, \sigma_j^2)^{R_{ij}} \prod_{k=1}^{K} f(W_{ik} | D_i = d, \boldsymbol{X}_i, \boldsymbol{\beta}_k^W)$$

Thus, the log-likelihood is as follows:

$$\log\left(L(\boldsymbol{\beta}^D, \boldsymbol{\beta}^R, \boldsymbol{\beta}^Y, \boldsymbol{\beta}^W, \sigma^2 | \boldsymbol{X}_i, \boldsymbol{Y}_i, \boldsymbol{W}_i)\right)$$

$$= \log\left(\sum_{d=0,1} P(D_i = d | \boldsymbol{\beta}^D, \boldsymbol{X}_i) \prod_{j=1}^{J} f\left(R_{ij} | D_i = d, \boldsymbol{X}_i, \boldsymbol{\beta}_j^R\right) f\left(Y_{ij} | D_i = d, \boldsymbol{X}_i, \boldsymbol{\beta}_j^Y, \sigma_j^2\right)^{R_{ij}} \prod_{k=1}^{K} f(W_{ik} | D_i = d, \boldsymbol{X}_i, \boldsymbol{\beta}_k^W)\right)$$

$$= \log\left(\sum_{d=0,1} \exp\left(\log\left(P(D_i = d | \boldsymbol{\beta}^D, \boldsymbol{X}_i) \prod_{j=1}^{J} f\left(R_{ij} | D_i = d, \boldsymbol{X}_i, \boldsymbol{\beta}_j^R\right) f\left(Y_{ij} | D_i = d, \boldsymbol{X}_i, \boldsymbol{\beta}_j^Y, \sigma_j^2\right)^{R_{ij}} \prod_{k=1}^{K} f(W_{ik} | D_i = d, \boldsymbol{X}_i, \boldsymbol{\beta}_k^W)\right)\right)\right)$$

$$= \log\left(\sum_{d=0,1} \exp\left(\log\left(P(D_i = d | \boldsymbol{\beta}^D, \boldsymbol{X}_i) \prod_{j=1}^{J} f\left(R_{ij} | D_i = d, \boldsymbol{X}_i, \boldsymbol{\beta}_j^R\right) f\left(Y_{ij} | D_i = d, \boldsymbol{X}_i, \boldsymbol{\beta}_j^Y, \sigma_j^2\right)^{R_{ij}} \prod_{k=1}^{K} f(W_{ik} | D_i = d, \boldsymbol{X}_i, \boldsymbol{\beta}_k^W)\right)\right)\right)$$

$$= \log sum \exp_{d=0,1} \log\left(P(D_i = d | \boldsymbol{\beta}^D, \boldsymbol{X}_i) \prod_{j=1}^{J} f\left(R_{ij} | D_i = d, \boldsymbol{X}_i, \boldsymbol{\beta}_j^R\right) f\left(Y_{ij} | D_i = d, \boldsymbol{X}_i, \boldsymbol{\beta}_j^Y, \sigma_j^2\right)^{R_{ij}} \prod_{k=1}^{K} f(W_{ik} | D_i = d, \boldsymbol{X}_i, \boldsymbol{\beta}_k^W)\right)$$

$$= \log\text{sum}\exp_{d=0,1}\left(\log P(D_i = d|\boldsymbol{\beta}^D, \boldsymbol{X}_i)\right.$$

$$+ \sum_{j=1}^{J}\log f\left(R_{ij}|D_i = d, \boldsymbol{X_i}, \boldsymbol{\beta}_j^R\right)$$

$$\left.+ \sum_{j=1}^{J}R_{ij}\log f\left(Y_{ij}|D_i = d, \boldsymbol{X_i}, \boldsymbol{\beta}_j^Y, \sigma_j^2\right) + \sum_{k=1}^{K}\log f(W_{ik}|D_i = d, \boldsymbol{X_i}, \beta_k^W)\right)$$

We use Stan's built-in log sum of exponentials function to define mixtures on the log scale, which improves numerical stability. This approach is recommended in the Stan manual for handling log-sum expressions in mixture models.

### Section S4. Derivation for our choice of informative priors

Binary variables were modeled as $W_k \sim Bernoulli(expit(\beta_0^k + \beta_1^k D))$, resulting in specificity $= 1 - expit(\beta_0^k)$ and sensitivity $= expit(\beta_0^k + \beta_1^k)$. We assign a prior of $\beta_l^k \sim Normal(\frac{\min(\beta_l^k) + \max(\beta_l^k)}{2}, 0.5)$ for $l = \{0,1\}$, where $\min(\beta_l^k)$ and $\max(\beta_l^k)$ refers to a prespecified restriction of each parameter's range which results in a restricted range for the sensitivity and specificity for correctly labeling T2-high asthma. For the parameters pertaining to positive skin prick test ($W_{PST}$), denoted as $\beta_0^{PST}$ and $\beta_1^{PST}$, we consider the fact that very few patients received an allergy skin prick test, resulting in only 1.5% having a positive result. Given that it is rare for a patient to receive an allergy skin prick test, fewer patients will have received this test relative to the presumed prevalence of T2-high asthma in our cohort. Thus, while we expect $P(D = 1|W_{PST} = 1)$ to be high, $W_{PST} = 0$ is uninformative, and there will be a high number of true and false negatives. As a result, we expect this indicator to have very high specificity but low sensitivity. We restrict the specificity to (0.99, 1) and sensitivity to (0.023, 0.038) by imposing the boundaries $\beta_0^{PST} \in (-9.2, -4.5)$ and $\beta_1^{PST} \in (-3.750 - \beta_0^{PST}, -3.245 - \beta_0^{PST})$.

### Section S5. Choice of priors used in asthma model

| Variable | Type | Model | Approach | Priors |
|---|---|---|---|---|
| Latent class | Binary | $D \sim Bern(expit(\beta_0 + \boldsymbol{\beta_X X}))$ | Weak priors | $\beta_0 \sim N(0,1)$ <br><br> $\beta_X^m \sim N(0,1)$ for $m = 1, \ldots, M$ |

| Average blood eosinophil count (EOS) | Continuous with missingness | $\text{std}(\log(Y_{EOS})) \sim N(\beta_0^{EOS} + \beta_1^{EOS} D, \sigma_{EOS})$ | Informative priors considering threshold Labelling at $0.15 \times 10^3$ cells/$\mu L$<br><br>Empirical Prevalence: 53.0%<br><br>Target Sensitivity: 75%<br><br>Target Specificity: 76% | $\beta_0^{EOS} \sim N(-0.68, 0.05)$<br><br>$\beta_1^{EOS} \sim N_{tr}(0.7 - \beta_0^{EOS}, 0.05, 0, \infty)$<br><br>$\sigma_{EOS} \sim N_{tr}(1, 0.1, 0, \infty)$ |
|---|---|---|---|---|
| Positive allergy skin prick test (PST) | Binary | $W_{PST} \sim \text{Bern}(\text{expit}(\beta_0^{PST} + \beta_1^{PST} D))$ | Informative priors<br><br>Empirical Prevalence: 1.5%<br><br>Target Sensitivity: (2.3%, 3.8%)<br><br>Target Specificity: (99%, 100%) | $\beta_0^{PST} \sim N_{tr}\left(\frac{-4.5 + (-9.2)}{2}, 0.5, -9.2, -4.5\right)$<br><br>$\beta_1^{PST} \sim N_{tr}(\frac{-3.245 + (-3.750) - 2\beta_0^{PST}}{2}, 0.5, -3.750 - \beta_0^{PST}, -3.245 - \beta_0^{PST})$ |
| Allergic ICD-10 code indicator (Allerg) | Binary | $W_{Allerg} \sim \text{Bern}(\text{expit}(\beta_0^{Allerg} + \beta_1^{Allerg} D)$ | Informative priors<br><br>Empirical Prevalence: 37.6%<br><br>Target Sensitivity: (52%, 94%)<br><br>Target Specificity: (83%, 100%) | $\beta_0^{Allerg} \sim N_{tr}\left(\frac{-1.61 + (-9.2)}{2}, 0.5, -9.2, -1.61\right)$<br><br>$\beta_1^{Allerg} \sim N_{tr}(\frac{0.068 + 2.769 - 2\beta_0^{Allerg}}{2}, 0.5, 0.068 - \beta_0^{Allerg}, 2.769 - \beta_0^{Allerg})$ |
| Total immunoglobulin E (IgE) | Continuous with missingness | $\text{std}(\log(Y_{IgE})) \sim N(\beta_0^{IgE} + \beta_1^{IgE} D, \sigma_{IgE})$ | Weak priors | $\beta_0^{IgE} \sim N(0, 1)$<br><br>$\beta_1^{IgE} \sim N(0, 1)$<br><br>$\sigma_{IGE} \sim N_{tr}(1, 0.1, 0, \infty)$ |
| Polygenic risk scores (SCORE) | Continuous with missingness | $\text{std}(Y_{SCORE}) \sim N(\beta_0^{SCORE} + \beta_1^{SCORE} D, \sigma_{SCORE})$ | Weak priors | $\beta_0^{SCORE} \sim N(0,1)$<br><br>$\beta_1^{SCORE} \sim N(0,1)$<br><br>$\sigma_{SCORE} \sim N_{tr}(1, 0.1, 0, \infty)$ |
| Pulmonary function test (PFT) | Continuous with missingness | $\text{std}(\log(Y_{PFT})) \sim N(\beta_0^{PFT} + \beta_1^{PFT} D, \sigma_{PFT})$ | Weak priors | $\beta_0^{PFT} \sim N(0,1)$<br><br>$\beta_1^{PFT} \sim N(0,1)$<br><br>$\sigma_{PFT} \sim N_{tr}(1, 0.1, 0, \infty)$ |
| Mean asthma encounters per year (enc) | Continuous with missingness | $\text{std}(\log(Y_{enc})) \sim N(\beta_0^{enc} + \beta_1^{enc} D, \sigma_{enc})$ | Weak priors | $\beta_0^{enc} \sim N(0,1)$<br><br>$\beta_1^{enc} \sim N(0,1)$<br><br>$\sigma_{enc} \sim N_{tr}(1, 0.1, 0, \infty)$ |
| Mean asthma exacerbations per year (exac) | Continuous with missingness | $\text{std}(\log(Y_{exac})) \sim N(\beta_0^{exac} + \beta_1^{exac} D, \sigma_{exac})$ | Weak priors | $\beta_0^{exac} \sim N(0,1)$<br><br>$\beta_1^{exac} \sim N(0,1)$<br><br>$\sigma_{exac} \sim N_{tr}(1, 0.1, 0, \infty)$ |
| Mean asthma emergency visits per year (ED) | Continuous with missingness | $\text{std}(\log(Y_{ED})) \sim N(\beta_0^{ED} + \beta_1^{ED} D, \sigma_{ED})$ | Weak priors | $\beta_0^{ED} \sim N(0,1)$<br><br>$\beta_1^{ED} \sim N(0,1)$<br><br>$\sigma_{ED} \sim N_{tr}(1, 0.1, 0, \infty)$ |
| Mean short-acting beta agonist prescriptions per year (SABA) | Continuous with missingness | $\text{std}(\log(Y_{SABA})) \sim N(\beta_0^{SABA} + \beta_1^{SABA} D, \sigma_{SABA})$ | Weak priors | $\beta_0^{SABA} \sim N(0,1)$<br><br>$\beta_1^{SABA} \sim N(0,1)$<br><br>$\sigma_{SABA} \sim N_{tr}(1, 0.1, 0, \infty)$ |
| Missingness in continuous variables | Binary | $R_j \sim \text{Bern}(\text{expit}(\beta_0^j + \beta_1^j D + \boldsymbol{\beta}_X^j \boldsymbol{X}))$ for<br>$j = \{EOS, IgE, SCORE, PFT, enc, exac, ED, SABA\}$ | Weak priors | $\beta_0^j \sim N(0,1)$<br><br>$\beta_1^j \sim N(0,1)$<br><br>$\beta_{X_m}^j \sim N(0,1)$ for $m = 1, \ldots, M$ |
| Record of inhaled corticosteroids prescription (ICS) | Binary | $W_{ICS} \sim \text{Bern}(\text{expit}(\beta_0^{ICS} + \beta_1^{ICS} D))$ | Weak priors<br><br>Empirical Prevalence: 36.2%<br><br>Target Sensitivity: (0%, 100%)<br><br>Target Specificity: (0%, 100%) | $\beta_0^{ICS} \sim N_{tr}(0, 0.5, -6, 6)$<br><br>$\beta_1^{ICS} \sim N_{tr}(-\beta_0^{ICS}, 0.5, -6 - \beta_0^{ICS}, 6 - \beta_0^{ICS})$ |
| Record of ICS and a long-acting beta-agonist | Binary | $W_{ICS/LABA} \sim$ | Weak priors | $\beta_0^{ICS/LABA} \sim N_{tr}(0, 0.5, -6, 6)$ |

| combination prescription (ICS/LABA) | | $Bern(expit(\beta_0^{ICS/LABA} + \beta_1^{ICS/LABA}D))$ | Empirical Prevalence: 58.0%<br><br>Target Sensitivity: (0%, 100%)<br><br>Target Specificity: (0%, 100%) | $\beta_1^{ICS/LABA} \sim N_{tr}(-\beta_0^{ICS/LABA}, 0.5, -6-\beta_0^{ICS/LABA}, 6-\beta_0^{ICS/LABA})$ |

### Section S6. Asthma model convergence assessment

Model convergence was assessed using the Rhat statistic and effective sample size (ESS) for each parameter. The scatter plot of Rhat values showed that they were all approximately 1.00 and are less than 1.01. Similarly, the scatterplot for ESS demonstrated all values were well above 1000 and their relative ESS exceeded 10% of the total 4000 post-warmup draws (above 400). These diagnostics indicate the chains mixed well and the posterior was adequately explored.

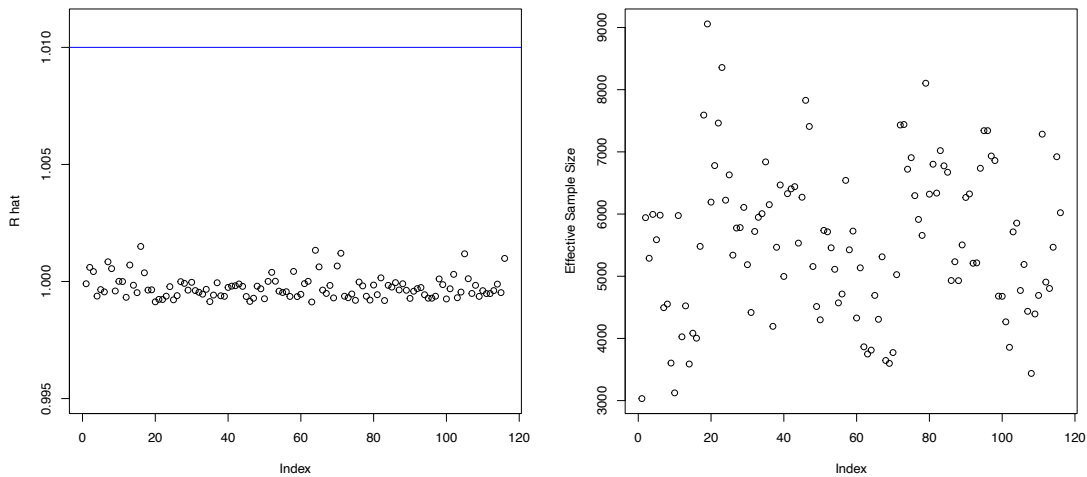

We display select trace plots to assess convergence. Below are the trace plots for the beta coefficients linking the latent variable to patient covariates. The four chains exhibit good mixing and converge rapidly to a common distribution, well before the designated burn-in period of 200 iterations. Trace plots for other parameters showed similar behavior.

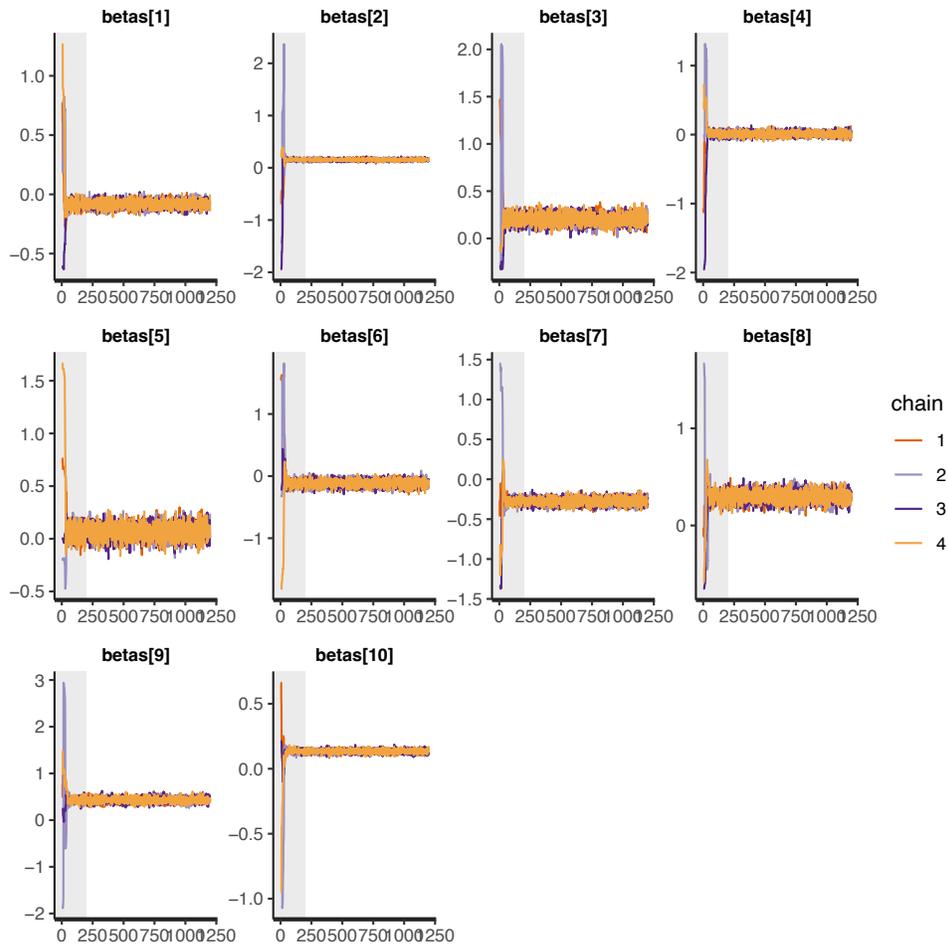